\documentclass[sigconf,screen,final]{acmart} 

\AtBeginDocument{%
  }

\setcopyright{acmlicensed}
\copyrightyear{2026}
\acmYear{2026}
\acmDOI{XXXXXXX.XXXXXXX}
\acmConference[NextAccel'26]{NextAccel'26:
International Workshop on Next-generation Accelerators For HPC, Data, and AI}{July 06, 2026}{Belfast, UK}
\acmISBN{978-1-4503-XXXX-X/2018/06}

\usepackage{booktabs}
\usepackage{tabularx}
\usepackage[table]{xcolor}

\usepackage{tikz}
\usepackage{standalone}
\usetikzlibrary{decorations.pathreplacing, calc, positioning, fit, backgrounds}

\begin{document}

\title{Post-Moore Technologies for Plasma Simulation: A Community Roadmap}


\settopmatter{authorsperrow=1}

\author{
Luca~Pennati*, Erik M.~Åsgrim*, Jeremy J.~Williams*, 
Stefan~Costea$^\dagger$,
David~Tskhakaya$^\P$,
Leon~Kos$^\dagger$,
Ales~Podolnik$^\P$,
Yi~Ju$^\|$,
Tapish~Narwal$^\diamond$,
Julian~Lenz$^\diamond$,
Michael~Bussmann$^\diamond$,
Urs~Ganse$^\bullet$,
Minna~Palmroth$^\bullet$,
Kallia~Chronaki$^\ddagger$,
Vassilis~Papaefstathiou$^\ddagger$,
Etienne~Renault$^\triangle$,
Felix~Jung$^\circ$,
Martin~Schulz$^\circ$,
Valentin~Seitz$^\S$,
Marta~Garcia-Gasulla$^\S$,
Filippo~Mantovani$^\S$,
Frank~Jenko$^\vardiamondsuit$,
Erwin~Laure$^\|$,
Stefano~Markidis*}


\affiliation{%
  \institution{*KTH Royal Institute of Technology, Stockholm, Sweden - \{pennati,erima,jjwil,markidis\}@kth.se}
  \institution{$^\dagger$University of Ljubljana, Ljubljana, Slovenia - \{stefan.costea,leon.kos\}@fs.uni-lj.si}
  \institution{$^\P$Institute of Plasma Physics of the CAS, Prague, Czech Republic - \{tskhakaya,podolnik\}@ipp.cas.cz}
  \institution{$^\|$Max Planck Computing and Data Facility, Garching, Germany - \{yi.ju,erwin.laure\}@mpcdf.mpg.de}
  \institution{$^\diamond$Helmholtz-Zentrum Dresden-Rossendorf, Dresden, Germany - \{t.narwal,j.lenz,m.bussmann\}@hzdr.de}
  \institution{$^\bullet$University of Helsinki, Helsinki, Finland - \{urs.ganse,minna.palmroth\}@helsinki.fi}
  \institution{$^\ddagger$Foundation for Research and Technology Hellas, Heraklion, Greece - \{khronaki,papaef\}@ics.forth.gr}
  \institution{$^\triangle$SiPearl, Maisons-Laffitte, France - etienne.renault@sipearl.com}
  \institution{$^\circ$Technical University Munich, Garching, Germany - \{felix.jung,martin.w.j.schulz\}@tum.de}
  \institution{$^\S$Barcelona Supercomputing Center, Barcelona, Spain - \{valentin.seitz,marta.garcia,filippo.mantovani\}@bsc.es}
  \institution{$^\vardiamondsuit$Max Planck Institute for Plasma Physics, Garching, Germany - frank.jenko@ipp.mpg.de}
  \country{}
}

\renewcommand{\shortauthors}{Pennati et al.}

\begin{abstract}
Plasma simulations are among the most computationally demanding scientific workloads, combining high-dimensional kinetic evolution, particle-mesh coupling, field solves, and data-intensive communication. As general-purpose processor scaling slows, post-Moore technologies are being explored to address bottlenecks in data movement, memory access, and power consumption. This paper provides a community perspective on the role of these technologies in plasma simulation, assessing three major classes: reconfigurable and data-path accelerators, non-von Neumann architectures, and quantum computing. Each is evaluated, in a co-design approach, against representative plasma workloads spanning particle-in-cell, continuum Vlasov, gyrokinetic, fluid/MHD, hybrid, and warm dense matter methods. We find that no single technology can replace existing HPC platforms. Instead, three tiers of opportunity emerge: FPGA-class and data-path accelerators offer near-term kernel offload and workflow-level data services, non-von Neumann architectures represent medium-term directions for operator-level acceleration, and quantum computing, although the least mature, is potentially the most disruptive for warm dense matter and inertial confinement fusion microphysics. We outline best practices for selective adoption and identify focused demonstrators, benchmarking, and modular software ecosystems as immediate community priorities.
\end{abstract}



\maketitle

\section{Introduction}
\label{sec:intro}
Plasma simulation plays a fundamental role in the analysis and prediction of a wide range of phenomena relevant to fusion energy~\cite{DiSiena2026gene_iter}, space plasmas~\cite{chen2017space_earth,lavorenti2023_space_mercury}, and high-energy-density physics~\cite{Couperus2017laser_picongpu}.  The underlying dynamics of plasma systems are inherently multiscale and non-equilibrium in nature. This requires modeling approaches that go beyond reduced fluid descriptions and instead retain kinetic effects, wave–particle interactions, and detailed phase-space dynamics. Because of this, plasma simulation codes and workflows rank among the most computationally demanding scientific applications.

In parallel with these scientific challenges, the HPC landscape is undergoing an important transformation. Although the transition to exascale systems has enabled substantial advances in simulation capability~\cite{markidis2025exascale,Wahlgren2025MI300}, it has also made evident that future performance gains cannot rely solely on conventional improvements in general-purpose processor technologies. Limitations related to data movement, memory access, synchronization, power consumption, and communication overhead are becoming dominant constraints in many scientific workloads. In this context, a new generation of emerging and post-exascale technologies is being explored as a means to complement established CPU- and GPU-based platforms and to address specific computational bottlenecks more effectively.

The goal of this paper is to provide a community perspective on the role of emerging post-Moore technologies in plasma simulations. More specifically, the report analyses the extent to which selected post-exascale technologies may offer tangible benefits for plasma applications. In this analysis, we consider their architectural characteristics, suitability for representative plasma workloads, and current technological maturity. The assessment is guided by a codesign perspective, in which the relevance of each technology is evaluated in relation to computational plasma physics methods.

\section{What Makes Plasma Simulation Hard?}
\label{sec:plasma_methods}
Plasma models form a hierarchy ranging from fluid/MHD descriptions to fully kinetic descriptions that evolve the distribution functions in phase space. Kinetic models are required when non-Maxwellian features, wave–particle interactions, collisionless dissipation, sheath physics, reconnection microphysics, or energetic particles play an essential role. At the most general level, collisionless kinetic physics is described by the Vlasov equation coupled to Poisson (electrostatic) or Maxwell (electromagnetic) field equations~\cite{Goldston1995plasma}. In practice, collisionality and sources are added through either Fokker–Planck/Landau-type operators or Monte Carlo collision models, depending on regime and solver class~\cite{Dimarco2015collisional,takizuka1977collision}. The numerical challenge is that the physics couples high-dimensional phase-space dynamics to lower-dimensional field equations. Nearly all kinetic solvers alternate between: (i) evolving particle or distribution degrees of freedom, and (ii) evaluating fields/moments, often by solving elliptic/hyperbolic PDEs. This coupling pattern is one reason kinetic codes are both compute- and data-intensive, and it motivates the later discussion on accelerators for particle–mesh coupling, sparse / structured PDE primitives, and in-path diagnostics. \\

\noindent\textbf{Particle-Based Kinetic Methods.}
Particle-in-Cell (PIC) methods represent the distribution function with macroparticles and couple them to mesh-based fields. Canonical PIC algorithms features, including particle shape functions, deposition/gather, stability considerations, numerical heating/noise control, and practical implementation details, are treated in standard references such as~\cite{BirdsallLangdon1985,HockneyEastwood1988}.

PIC is widely adopted because it avoids explicit velocity-space gridding and scales well to fully electromagnetic, relativistic, and complex-geometry settings. PIC exhibits a distinctive computational signature. Particle pushes are largely streaming, but charge and current deposition are scatter operations that involves fine-grained reductions and can be synchronization- and bandwidth-limited. Field updates may be stencil-like or solver-dominated depending on formulation~\cite{pennati2025poisson}. In Plasma-PEPSC, BIT1~\cite{Tskhakaya2010BIT1,williams2023leveraging,williams2025BIT1,williams2026multigpu} and PIConGPU~\cite{bussmann2013picongpu} are examples of PIC solvers. \\

\noindent\textbf{Continuum Eulerian Kinetic Solvers.}
Continuum kinetic methods discretize the distribution function f(x,v,t) directly on a phase-space mesh using finite-volume, finite-difference, semi-Lagrangian, or finite-element and discontinuous-Galerkin schemes~\cite{cheng1976vlasov,Palmroth2018vlasov}. These methods avoid particle noise and can reach very high accuracy, but they face the “curse of dimensionality” since memory and computation scale steeply with phase-space dimension~\cite{aasgrim2026tensor}. These methods represent the full distribution continuously, thus, they are particularly valuable for resolving weak instabilities, fine velocity-space filamentation, trapping structures, or distribution tails without statistical sampling error.

Their dominant numerical kernels are phase-space advection, interpolation/reconstruction, and repeated computation of moments coupled to Poisson or Maxwell solves. As a result, they are often limited by memory capacity and data movement rather than raw floating-point throughput alone. In practice, continuum Vlasov solvers often exploit reduced dimensionality (e.g., 1D1V, 2D3V), adaptivity, and sparse representations to make high-fidelity kinetic calculations tractable. The Plasma-PEPSC Vlasiator code~\cite{ganse2023vlasiator} uses a Vlasov solver to model the dynamics of ions. \\

\noindent\textbf{Gyrokinetics and reduced kinetic models.}
For strongly magnetized plasmas, gyrokinetic theory reduces the 6D kinetic problem by averaging over fast gyromotion, producing a 5D description that retains the essential low-frequency turbulence and wave–particle physics central to magnetic confinement fusion~\cite{FriemanChen1982,BrizardHahm2007}. The reduction is justified when the dynamics of interest are slow compared with the cyclotron motion and anisotropic with respect to the magnetic field, which is precisely the regime relevant to core and edge turbulence studies in tokamaks and stellarators.

Gyrokinetic models can be solved with either particle-based or continuum discretizations, and they typically couple kinetic evolution along magnetic field lines to reduced electrostatic or electromagnetic field equations. Compared with full Vlasov-Maxwell models, they deliver a substantial reduction in cost while preserving the microphysics most relevant for turbulent transport, zonal flows, and wave-particle interactions in magnetized fusion plasmas. They remain computationally demanding, however, because the problem is still five-dimensional and often involves complex magnetic geometry, stiff parallel dynamics, and collision operators. In Plasma-PEPSC, the GENE code~\cite{jenko2000gene,Germaschewski2021gene,merlo2021first,di2026full} is an example of gyrokinetic code. \\

\noindent\textbf{Hybrid and Coupled Plasma Workflows.}
Many important plasma problems are intrinsically multiscale. A fluid description may be adequate over much of the domain, while localized regions require kinetic fidelity (e.g., reconnection layers, sheaths, shocks). This motivates hybrid models and embedded kinetic regions, where kinetic solvers are coupled to fluid/MHD solvers. A well-known example is MHD-EPIC~\cite{Daldorff2014MHDEPIC}, which couples a Hall-MHD model to an implicit PIC model with two-way coupling and matched resolution near the coupling interface. In Plasma-PEPSC, the Vlasiator is an example of hybrid code, where the ion dynamics is solved kinetically and electron dynamics is solved with fluid modeling. \\

\noindent\textbf{Warm dense matter and molecular dynamics.}
Warm dense matter (WDM) occupies a distinctive regime in which strong coupling, quantum effects, and electronic excitation are all simultaneously important, making it less suitable for the standard approximations used in either classical plasma or ground-state materials simulations~\cite{vorberger2025wdm_roadmap}. Its main first-principles techniques are Kohn–Sham density functional theory (DFT), DFT-based molecular dynamics~\cite{Bonitz2020_wdm_abinitio} and path-integral Monte Carlo (PIMC)~\cite{ceperley1995_pimc}. Computationally, these methods are dominated by self-consistent electronic-structure kernels rather than by particle–mesh coupling: Kohn–Sham workflows are driven by FFTs, iterative diagonalization, matrix operations, and communication. PIMC is dominated by stochastic path sampling, action evaluation, and statistical reductions. WDM therefore forms a distinct workload class centered on transforms, linear algebra, Monte Carlo sampling, and communication-heavy solves.

\subsection{Workload Characteristics Across Plasma Use Cases}
The dominant workload patterns in plasma simulations are determined primarily by the numerical method rather than by the application itself. PIC and related particle methods expose substantial streaming parallelism in the particle push, but are typically constrained by irregular memory access, fine-grained reductions, and synchronization. Implicit formulations also involve solving linear or nonlinear systems, introducing sparse linear algebra and iterative solver workloads. Continuum kinetic and Vlasov solvers replace particle irregularity with large multidimensional array sweeps, interpolation and reconstruction, remapping, moment evaluation, and field-coupled updates, making them primarily bandwidth and data-movement-dominated workloads. Gyrokinetic solvers retain a similarly high-dimensional array structure, but their computational profile more often combines tensor-like updates, reductions over velocity space, collision operators, spectral transforms, and field solves. Fluid and MHD solvers are instead dominated by regular stencil or flux computations, local reconstructions, and sparse linear solves. Across all of these techniques, data communication plays a central role in distributed-memory execution, through halo exchanges, particle migration, global reductions, and solver communication, affecting the scalability at large machine size. 
We summarize in Table~\ref{tab:workloads} the main workloads that characterize different simulation methods and plasma domain fields.

\begin{table*}[t!]
    \centering
    \caption{Workload characteristics across different simulation methods and use cases.
    \checkmark\ denotes a dominant workload, $\circ$\ a recurring but secondary workload, and - a generally minor or absent one.}
    \renewcommand{\arraystretch}{1.1}
    \begin{tabular}{l|ccccccccc}
        \toprule
        Case/Method
        & Streaming
        & \shortstack{Interpolations}
        & \shortstack{Reductions}
        & \shortstack{Stencil\\Calc.}
        & \shortstack{Linear\\Algebra}
        & \shortstack{FFT / Spectral\\sol.}
        & Sorting
        & \shortstack{MC\\Sampling}
        & \shortstack{Communication} \\
        \midrule
        PIC                      & \checkmark & \checkmark & \checkmark & \checkmark      & $\circ$      & --          & $\circ$      & --         & \checkmark \\
        Kinetic Vlasov           & \checkmark & \checkmark & \checkmark & $\circ$      & $\circ$      & $\circ$       & $\circ$      & --         & \checkmark \\
        Gyrokinetic              & $\circ$      & $\circ$      & \checkmark & $\circ$      & $\circ$      & \checkmark  & --         & --         & \checkmark \\
        Fluid / MHD              & \checkmark & $\circ$      & $\circ$      & \checkmark & $\circ$      & $\circ$       & --         & --         & \checkmark \\
        WDM (DFT)    & $\circ$      & --         & $\circ$      & --         & \checkmark & \checkmark  & --         & --         & \checkmark \\
        WDM (PIMC)               & --         & --         & \checkmark & --         & --         & --          & --         & \checkmark & \checkmark \\
        \bottomrule
    \end{tabular}
    \label{tab:workloads}
\end{table*}

\section{Near-Term Opportunities: Reconfigurable and Data-Path Accelerators}
\label{sec:fpgas}

\noindent\textbf{Reconfigurable accelerators.}
Field-Programmable Gate Arrays (FPGAs)~\cite{bobda2022fpgas} and Coarse-Grained Reconfigurable Architectures (CGRAs)~\cite{liu2019cgras} are well-established technologies that enable computations to be performed \emph{as data flows through the hardware}. They have already found applications for diagnostic and control workflows in experimental setups~\cite{Hanson2009DigitalControl,Carvalho2010ISTTOKTomographyFPGA,Vega2018FPGA_DisruptionPredictor,Kolasinski2023diagnosticWEST}.

In the context of simulations, the most mature FPGA targets are sub-kernels within particle-based algorithms that can be reformulated in a dataflow-oriented manner. Across the literature, two implementation strategies recur: \emph{(i)} spatial tiling, in which particles are sorted into tiles or cells to preserve locality of updates, and \emph{(ii)} buffer replication and conflict management during deposition, which mitigate race conditions without compromising pipeline throughput.
An example is the FPGA electrostatic PIC implementation by Almomany et al., which targets the computationally intensive particle-to-interpolation phase and discusses architecture choices that reduce memory access latency on an FPGA platform~\cite{Almomany2024FPGA_ElectrostaticPIC}. A complementary approach is given by Guidotti’s paper on accelerating particle–mesh algorithms using FPGAs and OmpSs@OpenCL~\cite{Guidotti2025ParticleMeshFPGAs}. The thesis emphasizes strict particle tiling and explicitly managed local memory, and it introduces multiple copies of current buffers to resolve dependencies during deposition. 

Beyond single-kernel studies, the Full-PIC Hall thruster acceleration work by Sakai et al.~\cite{Sakai2016FullPIC_CPUFPGA} illustrates a realistic hybrid approach. This retains the full simulation code on the host, but offloads heavy steps like particle-grid interpolation to FPGA logic using HLS and pipelined stencil/dataflow structures, achieving large kernel speedups, up to $\sim10\times$. This paper is a proof-of-principle that meaningful gains can be achieved when a workload segment is sufficiently regular to be hardened into a streaming FPGA pipeline.\\

\noindent\textbf{Data-path accelerators.} 
In large kinetic simulations, many stages can be limited by data movement, thus making custom accelerators, such as DPUs/SmartNICs, particularly relevant since they allow parts of the data-management workload (filtering, partitioning, compression) to run off-path from the host and, in some cases, in parallel with the primary simulation.

A closely related example for plasma-style particle outputs is provided by Processing Particle Data Flows with SmartNICs~\cite{Liu2022SmartNICParticleFlows}. This work implements a partitioning algorithm for particle datasets on an NVIDIA BlueField-2 SmartNIC and analyzes how on-card compression/decompression hardware can affect the performance of in-transit workflows, where data are unpacked, processed, and repacked.
Additional work has shown that SmartNICs can offload data-service tasks from the host. In particular, BlueField-2 devices can be used to sort, partition, redistribute, and reorganize particle data into forms better suited to downstream analysis~\cite{ulmer2023smartnic_particlesifting}. More general studies~\cite{Wahlgren2024smartnic,wahlgren2026buddy} have further demonstrated the potential of SmartNICs to improve data movement, communication, caching, and prefetching by offloading selected data-transfer and communication tasks from the host.

DPUs have also been explored as accelerators for linear algebra pipelines, which are common across all major plasma simulation methods. A representative example is provided by Suresh et al.~\cite{Suresh2024BlueField3_vecOp_krylov}, who accelerated a Krylov-based linear solver by offloading matrix multiplication operations to a BlueField-3 SmartNIC, reporting speedups of up to 20\%. The best fit targets in plasma simulations for FPGA-class and data-path accelerators are summarized in Table~\ref{tab:tf1_technology_summary}.

\begin{table*}[h!]
\centering
\caption{Comparison of FPGA-class and data-path accelerator technologies for plasma workflows.}
\footnotesize
\setlength{\tabcolsep}{4pt}
\renewcommand{\arraystretch}{1.12}
\begin{tabularx}{\textwidth}{>{\raggedright\arraybackslash}p{2.5cm}>{\raggedright\arraybackslash}p{4.0cm}>{\raggedright\arraybackslash}p{4.0cm}>{\raggedright\arraybackslash}X}
\toprule
\textbf{Technology} & \textbf{Key strength} & \textbf{Best fit} & \textbf{Readiness in plasma workflows} \\
\midrule
FPGA / CGRA & 
Custom or semi-structured dataflow execution, explicit buffering, deterministic latency, and strong locality control for regular kernels. & 
Streamable particle-mesh kernels, regular loop/dataflow computations, and latency-sensitive diagnostics or control tasks. & 
Medium for FPGA-based simulation kernels and high for diagnostics/control; medium-low overall for CGRA-style deployment, with limited plasma-specific evidence so far~\cite{Almomany2024FPGA_ElectrostaticPIC,Sakai2016FullPIC_CPUFPGA,Hanson2009DigitalControl,Carvalho2010ISTTOKTomographyFPGA,Vega2018FPGA_DisruptionPredictor,Kolasinski2023diagnosticWEST,liu2019cgras}. \\
\midrule
SmartNIC & 
In/near-network filtering, partitioning, repacking, and lightweight analytics close to the communication path. & 
Particle/field stream preprocessing, workflow staging, and data-movement reduction. & 
Medium for data-path tasks; low-medium for plasma-specific adoption~\cite{wei2023smartNIC,Liu2022SmartNICParticleFlows,kfoury2024smartnic}. \\
\midrule
DPU & 
Infrastructure-side compute with embedded cores, DMA, memory services, and engines such as compression. & 
Compression, storage and network services, workflow staging, and host-offload tasks. & 
High for infrastructure offload; medium for integration into plasma workflows~\cite{tibbets2026dpu_smartnic,Li2024BlueFieldCompression}. \\
\bottomrule
\end{tabularx}
\label{tab:tf1_technology_summary}
\end{table*}

\section{Medium-Term Opportunities: Non-von-Neumann Architectures}
\label{sec:non_vonneumann}

Most present-day HPC systems remain based on the von Neumann model, in which memory and compute/control units are physically separated, making data movement a major source of cost. In this context, \emph{non-von Neumann architectures} are of interest because they change the execution model itself rather than simply accelerating conventional instruction streams, for example, through event-driven updates, in-memory algebra, or analog physical computation with the goal of improving time-to-solution and/or energy-to-solution. \emph{Non-von Neumann architectures} comprise a broad range of different technologies, here we focus on the following, most established architectures: \emph{(i)} neuromorphic computing;
\emph{(ii)} thermodynamic computing; \emph{(iii)} compute-in-memory and analog computing; \emph{(iv)} photonic computing; and \emph{(v)} domain-specific processors.
We summarize the main characteristics and applications in plasma workflow for each technology class in Table~\ref{tab:nonvonneumann_summary}. \\

\noindent\textbf{Neuromorphic computing.}
Neuromorphic processors distribute memory and computation across neuron- and synapse-like elements that communicate through sparse spike events rather than through conventional dense instruction streams~\cite{mead1990neuromorphic,davies2021neuromorphic,Benjamin2014neurogrid,moradi2018dynaps}. Their main advantage is the low-power execution of sparse, iterative, event-driven, or latency-constrained tasks. Thus, it should be seen not as a generic replacement for conventional plasma HPC capable of providing dense floating-point throughput.

The most promising plasma-relevant uses are in selected PDE, stochastic kernels, and reduced models. Recent work has shown that spiking neuromorphic hardware can execute FEM-based Poisson-type solves, including irregular meshes, on Loihi~2~\cite{TheilmanAimone2025NeuroFEM}. This is relevant to electrostatic field solves and related elliptic subproblems. Neuromorphic implementations of random-walk and diffusion-type methods have also been demonstrated on TrueNorth and Loihi hardware, suggesting applicability to stochastic transport and Monte Carlo-like formulations~\cite{Smith2022neuromorphic_ranodm_walk,smith2020neuromorphic_pde}. In addition, spiking systems are being explored for physics-informed learning, reduced dynamical models, and learned time integration for ODEs and PDEs~\cite{TheilmanEtAl2024SpikingPINNsLoihi2,ZhangEtAl2022SpikingMarchingScheme_arXiv}.

\noindent\textbf{Thermodynamic Computing.}
Thermodynamic computing is a stochastic, physics-based computing paradigm that uses noisy dissipative physical systems as computational backends. By tuning the system energy landscape and couplings, its equilibrium statistics can encode a target computation~\cite{conte2019thermodynamic}.

The most plasma-relevant applications are field solves, inference, and reduced-model kernels. Thermodynamic linear algebra and stochastic-circuit hardware have demonstrated Gaussian sampling and matrix inversion~\cite{Aifer2024thermodynamic_linearalgebra,Melanson2025thermodynamic_ai}, which are relevant to Poisson-like solves, implicit updates, and preconditioning. Thermodynamic matrix exponentials~\cite{Duffield2025thermodynamics_matrix_exp} are also promising because exponential operators appear in stiff plasma time integrators, including resistive MHD and drift-kinetic models~\cite{einkemmer2017exponential_integration_mhd,Crouseilles2018exponentia_integrators_dk}. Another relevant direction is Bayesian plasma analysis, where thermodynamic Bayesian inference can support posterior sampling and uncertainty quantification~\cite{aifer2024thermodynamic_bayesian,Kruger2024bayesian_plasma}. These primitives may also benefit surrogate and reduced-order plasma models involving covariance operations, sampling-based calibration, and uncertainty-aware inference. A narrower but potentially useful application is Monte Carlo-related plasma workflows. Since thermodynamic devices are naturally matched to Langevin-style and posterior sampling, they may support Markov Chain Monte Carlo tasks such as proposal generation and posterior exploration in Bayesian plasma-profile inference~\cite{Nishizawa2025mcmc_eq_reconstruction}.

At present, these opportunities remain prospective, as existing devices are still small-scale and constrained by precision, calibration, and embedding limitations~\cite{Melanson2025thermodynamic_ai,aifer2024error_mitigation_therm}. Recent error-mitigation results suggest that correction strategies can partly compensate for low-precision thermodynamic hardware~\cite{aifer2024error_mitigation_therm}, making such devices plausible for reduced models and moderate-accuracy plasma subroutines.

\noindent\textbf{Compute-in-memory.} Compute-in-memory (CIM) moves arithmetic into, or immediately next to, the memory array in order to reduce data movement~\cite{Zidan2018CIM_memristive,Singh2025aimc}. In digital CIM, arithmetic or logic is embedded in memory macros using conventional digital circuits, while in analog in-memory computing (AIMC), matrix-vector products are executed directly in the memory fabric, for example in resistive or memristive crossbars. Matrix coefficients are encoded as device conductances, input values are applied as voltages or currents, and the summed output currents realize weighted accumulation through Ohm's and Kirchhoff's laws. 

Natural targets for CIM and AIMC are sparse and repeated linear-algebra operations arising in linear systems. Therefore, the main opportunities for plasma simulations are in Poisson-like solves, implicit field updates, fluid-model linear systems, preconditioners, and surrogate-model inference. A particularly relevant example is the memristor-crossbar PDE solver of Zidan et al., which was integrated into a plasma workflow by replacing the Poisson step in an argon inductively coupled plasma reactor model~\cite{ZidanEtAl2018MemristorPDESolver}. More broadly, CIM/AIMC has already been used for linear-system solution, sparse matrix operations, and PDE-oriented workloads, including mixed-precision iterative refinement and sparse matrix-vector products~\cite{LeGallo2018MixedPrecision,Li2023cim_sparse_matrix,Zuo2025cim_matrix_solve,Li2025cim_matrixIteration,Qi2023cim_pde}. The main limitation is precision: device variability, noise, converter resolution, and drift remain critical issues for scientific computing. However, bit slicing, residue methods, calibration, and mixed-precision refinement provide possible mitigation routes~\cite{song2024cim_highprecision,Mannocci2026CIM_highprecision}. Accordingly, CIM/AIMC is best seen as a candidate accelerator for repeated algebraic subroutines with controllable accuracy requirements, rather than a replacement for end-to-end simulation pipelines. 

\noindent\textbf{Photonic computing.}
Photonic computing uses optical propagation, interference, modulation, and detection to implement numerical operations, especially linear transforms (e.g., Fourier transforms), matrix operations, and neural-network inference, with data encoded in the amplitude, phase, wavelength, or polarization of light~\cite{Zhou2022photonic_matrixmul,Hua2025photonics,Zhang2025photonic_nn}. Its main attraction is the combination of very high bandwidth, low propagation latency, and intrinsic parallelism across wavelengths and modes, which makes it particularly appealing for analog linear algebra and transform-heavy workloads.

For plasma workflows, the most plausible role of photonic hardware is the acceleration of selected linear-algebra kernels, PDE-related operators, and possibly reduced neural-network surrogates. Recent demonstrations include photonic solvers for Laplace, Poisson, and diffusion-type equations~\cite{Yuan2025micromb_pde,Ye2024photonic_recon_pde,Tang2025photonic_pde_neural}, as well as reconfigurable photonic iterative processors for matrix inversion and related linear-algebra tasks~\cite{Chen2024photonic_iterative}. In addition, optical neural-network architectures have been proposed for fast operator evaluation and inference, demonstrating time-dependent and time-independent PDE solving across several scientific domains, including magnetostatic Poisson problems, Navier-Stokes dynamics, Maxwell equations, and coupled multiphysics systems~\cite{Tang2025photonic_pde_neural}. However, photonic computing has not yet been integrated into plasma workflows, and broader adoption still depends on progress in scalability, calibration and control complexity, nonlinear-function support and precision management~\cite{Kutluyarov2023neuromorphic_photonic,Gupta2025neuromorphic_photonic,Zhang2025photonic_nn}. Thus, its near/medium-term relevance is as a specialized accelerator for a narrow class of linear-operator dominated kernels and inference-oriented surrogates. \\

\noindent\textbf{Domain-specific many-body processors.}
Domain-specific many-body processors are accelerators for particle-based simulations that hardwire the dominant interaction kernels into dedicated pipelines, while a host processor handles integration, control, and communication~\cite{Makino1997grape4,Ohmura2014mdgrape4,shaw2008Anton,shaw2014anton2}. When a single interaction pattern dominates the workload, substantial gains can be achieved by co-designing the hardware around that kernel.

This model is especially relevant for strongly coupled plasma molecular dynamics, warm dense matter, and other microphysics workflows dominated by Coulomb or short-range particle interactions. The GRAPE family demonstrated this approach for direct $N$-body force evaluation~\cite{Makino1997grape4}, while MDGRAPE extended it to molecular dynamics and non-bonded interactions, including system-on-chip designs and support for scalable execution~\cite{Susukita2003mdgrape2,Ohmura2014mdgrape4}. MDGRAPE-4A is particularly relevant because it couples dedicated interaction hardware with particle-grid and grid-grid operations for long-range electrostatics~\cite{morimoto2021mdgrape4a}. Anton, although developed for biomolecular simulation, further illustrates the same architectural lesson at larger scale~\cite{shaw2008Anton,shaw2014anton2}. For plasma science, these systems suggest that dedicated accelerators may be worthwhile when particle-particle or particle-mesh interaction kernels dominate the total simulation cost.

\begin{table*}[t!]
\centering
\caption{Comparison of non-von Neumann architecture families for plasma workflows.}
\footnotesize
\setlength{\tabcolsep}{4pt}
\renewcommand{\arraystretch}{1.12}
\begin{tabularx}{\textwidth}{>{\raggedright\arraybackslash}p{2.5cm}>{\raggedright\arraybackslash}p{4.0cm}>{\raggedright\arraybackslash}p{4.0cm}>{\raggedright\arraybackslash}X}
\toprule
\textbf{Technology} & \textbf{Key strength} & \textbf{Best fit} & \textbf{Readiness in plasma workflows} \\
\midrule
Neuromorphic comp. &
Event-driven, sparse communication, low power, distributed local state. &
Sparse PDE/linear-system kernels; spiking surrogates and time integrators; event-driven control and sensing. &
Low-medium for PDE kernels~\cite{TheilmanAimone2025NeuroFEM,smith2020neuromorphic_pde}; low-medium for surrogate/SciML~\cite{TheilmanEtAl2024SpikingPINNsLoihi2}; medium for control/sensing~\cite{Jansen2025neuromorphic_controller,jian2025neuromorphicDVS_1}. \\
\midrule
Thermodynamic comp. &
Physics-based stochastic computation; natural fit for sampling, matrix functions, and probabilistic inference. &
Structured linear algebra; matrix inverses/determinants; matrix exponentials; Bayesian inference and uncertainty-aware surrogate training. &
Low; scientific computing demonstrations exist~\cite{Aifer2024thermodynamic_linearalgebra,Melanson2025thermodynamic_ai}, but the plasma relevance is still prospective. \\
\midrule
CIM/AIMC &
In-memory matrix-vector multiplication; reduced data movement. &
Poisson-like linear algebra in reduced models; preconditioner-level kernels; data-analysis pipelines. &
Low-medium; one plasma-specific demonstration~\cite{ZidanEtAl2018MemristorPDESolver}; broader scientific-computing evidence growing~\cite{Li2023cim_sparse_matrix,Zuo2025cim_matrix_solve}. \\
\midrule
Photonic comp. &
Ultralow-latency, high-bandwidth analog linear algebra; wavelength parallelism. &
Structured PDE operators (Poisson, Laplace); matrix-MAC acceleration; iterative linear algebra. &
Low; scientific-computing demonstrations exist~\cite{Yuan2025micromb_pde,Hua2025photonics} but no plasma-specific deployment yet. \\
\midrule
Many-body proc. &
Dedicated interaction pipelines; fixed data paths; high arithmetic density. &
Coulomb-dominated force evaluation; particle-mesh electrostatics; WDM/strongly coupled plasma MD. &
High in astrophysics and biomolecular MD~\cite{Makino1997grape4,Ohmura2014mdgrape4,shaw2008Anton}; no plasma-specific adoption so far. \\
\bottomrule
\end{tabularx}
\label{tab:nonvonneumann_summary}
\end{table*}

\section{Long-Horizon Opportunities: Quantum Computing for Plasma Science}
\label{sec:quantum}
Quantum computing (QC) is an emerging computational paradigm that exploits quantum dynamics in ways fundamentally different from classical computing. Although current quantum computers remain limited in scale and noise levels, rapid progress in hardware and algorithm development continues to motivate scientific use cases. For plasma science in particular, work to date has mainly focused on reduced demonstrator problems on small devices or simulators. In the longer term, potential relevance may extend to subproblems with quantum many-body structure or efficient Hamiltonian encodings.

Two application families appear especially relevant for future use of quantum computing in plasma simulation. The first is kinetic plasma dynamics and wave-particle physics that can be mapped to Hamiltonian evolution in linearized or hybrid formulations, including Vlasov(-Poisson/-Maxwell) test problems and linear response regimes~\cite{Engel2019VlasovPRA}. The second is warm dense matter (WDM) and inertial confinement fusion (ICF) microphysics, where stopping power, opacity, equation of state (EOS), and finite-temperature response depend on strongly coupled electron-ion dynamics that remain difficult to validate with classical methods alone~\cite{vorberger2025roadmap,Rubin2024StoppingPowerPNAS}.

\begin{figure}[h]
    \centering
    \includegraphics[width=0.72\linewidth]{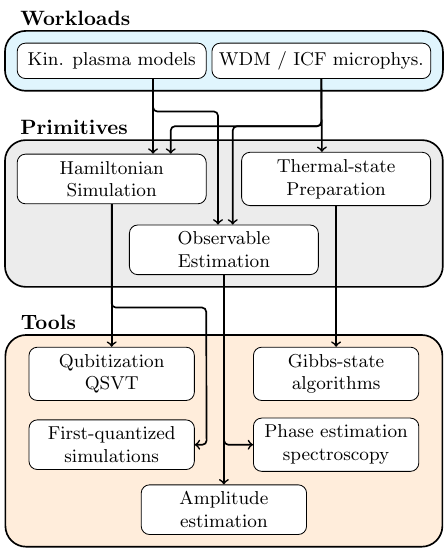}
    \caption{Representative mapping between plasma/WDM workloads and the quantum algorithmic primitives commonly used to address them.}
    \label{fig:quantum_diagram}
\end{figure}

Across the literature, a few algorithmic frameworks and computational primitives frequently appear in quantum-simulation proposals relevant to plasma physics. These include Hamiltonian simulation, thermal-state preparation, and observable estimation. For Hamiltonian simulation, modern approaches typically combine block encoding, qubitization, and quantum singular value transformation (QSVT)~\cite{low_hamiltonian_2019,gilyen_quantum_2019}. This combination offers near-optimal scaling in simulation time and precision, while supporting matrix-function operations central to plasma simulation. Closely related first-quantized simulation frameworks and resource analyses provide explicit qubit and gate counts for many-body problems and are often used as technical baselines for plasma-relevant microphysics estimates~\cite{su_fault-tolerant_2021}. For finite-temperature regimes, thermal-state preparation and Gibbs-state algorithms provide routes to thermodynamic properties and response quantities, with recent work and complexity analyses identifying regimes where efficient thermalization is plausible~\cite{chen_efficient_2025,rouze_efficient_2025,rouze_optimal_2026}. Observable-estimation methods complement these primitives by targeting measurable quantities directly through phase-estimation, amplitude-estimation, and sampling-based Krylov protocols~\cite{abrams_quantum_1999,montanaro_quantum_2015,yu_quantum-centric_2025}.

For kinetic plasma dynamics, several recent studies show how linearized or discretized models can be reformulated as Hamiltonian-simulation problems. A foundational plasma-specific result shows that linearizing Vlasov-Maxwell around a Maxwellian background yields a Hamiltonian structure suitable for quantum simulation, with analysis in the electrostatic Landau-damping limit~\cite{Engel2019VlasovPRA}. A hybrid quantum-classical Vlasov-Maxwell approach has also been demonstrated, where the Vlasov step is implemented through QSVT-based Hamiltonian simulation and coupled to a classical Maxwell solver, including 1D advection and 1D1V two-stream tests~\cite{Higuchi2025JPP_HS}. Two additional studies extend this beyond the purely collisionless limit. One formulates a quantum algorithm for the linear Vlasov equation with collisions using a Fourier-Hermite representation, combining Hamiltonian simulation for collisionless evolution with quantum ODE solvers for collisional dynamics~\cite{Ameri2023VlasovCollisionsPRA}. Another work analyzes QSVT-based Hamiltonian-simulation complexity for linearized Vlasov-Poisson in the Landau-damping regime and discusses the role of amplitude amplification in resource tradeoffs~\cite{Toyoizumi2024QSVT_HS_PRA}. Taken together, these studies suggest a hierarchy from reduced linearized demonstrators to hybrid and extended-physics formulations, while leaving open issues in discretization, scaling, and non-Hamiltonian effects.

For WDM and ICF microphysics, current work often focuses on observables directly relevant to model validation and performance assessment. Target design and performance assessment depend on stopping power, opacity, EOS, and finite-temperature response quantities that are difficult to compute with classical methods in regimes where strong correlation, degeneracy, and thermal effects compete~\cite{vorberger2025roadmap}. In this context, thermal-state preparation and Gibbs-state algorithms are commonly treated as enabling capabilities. Recent work proposes an efficient quantum thermal simulation framework with a detailed-balance structure resembling Markov chain Monte Carlo~\cite{chen_efficient_2025}. The method uses a Lindbladian quantum Markov process that preserves locality and enables Gibbs preparation under suitable mixing conditions. Complementary complexity analyses identify regimes of high-temperature rapid mixing and clarify conditions under which quantum Gibbs samplers can thermalize efficiently~\cite{rouze_efficient_2025,rouze_optimal_2026}.

Plasma-relevant outputs are often obtained through targeted measurement routines rather than full-state reconstruction. Observable-estimation tools therefore appear frequently in realistic quantum workflows. Phase-estimation spectroscopy methods provide one route to extracting spectral information~\cite{abrams_quantum_1999}, while amplitude-estimation algorithms provide the standard pathway to accelerating Monte Carlo-type sampling of expectation values~\cite{montanaro_quantum_2015}. More recently, quantum sampling-based Krylov approaches have been explored as a hybrid route to approximate eigenvalues and response properties from samples of time-evolved states, potentially providing a more hardware-efficient route to spectral information in near-term devices.
~\cite{yu_quantum-centric_2025}. A particularly direct plasma microphysics example is a first-quantized stopping-power protocol that computes projectile energy loss from coupled electron-projectile dynamics and provides explicit fault-tolerant estimates on the order of $10^3$ logical qubits and $10^{15}$-$10^{17}$ Toffoli gates for fully converged ICF-relevant cases~\cite{Rubin2024StoppingPowerPNAS}.

Additional works have explored hybrid classical-quantum workflows for scientific simulations. Hegde et al.~\cite{hegde2026hybrid_pic} developed an electrostatic PIC where the conventional Poisson solver is replaced by a hybrid classical-quantum neural network. Markidis et al.~\cite{markidis2026qpu_microkernels} proposed a general framework for solving partial differential equations with a Monte Carlo approach, where the quantum hardware is used as an accelerated sampler.

Overall, the literature indicates that quantum-computing studies in plasma science are currently focused on reduced-kinetic and microphysics demonstrators, with an emphasis on mapping workloads to core primitives and quantifying resource and measurement requirements. Figure~\ref{fig:quantum_diagram} provides an illustrative, non-exhaustive summary of these mappings.

\section{Implications for the Plasma Computing Community}
\label{sec:best_practices}

The assessment presented in the previous sections leads to several implications for the plasma computing community. 
\begin{enumerate}
    \item No single post-Moore technology is currently positioned to replace conventional CPU- and GPU-based HPC platforms for production plasma simulation. Instead, the most credible path is selective, codesign-driven adoption, in which emerging accelerators are integrated into existing workflows at points where their architectural strengths align with specific computational bottlenecks. 
    \item The technologies reviewed span a wide range of maturity: FPGA-class and data-path accelerators are ready for targeted deployment in near-term workflows, non-von Neumann architectures offer medium-term research opportunities for selected kernels and surrogate models, while quantum computing remains a long-horizon but potentially disruptive direction for problems with intrinsic quantum many-body structure.
    \item Effective adoption will require the plasma community to invest in modular software architectures, portable kernel interfaces, and structured benchmarking efforts that enable seamless integration of emerging hardware into current pipelines, with evaluation under realistic scientific conditions rather than on isolated micro-benchmarks alone.
\end{enumerate}
Table~\ref{tab:best_practice} summarizes the readiness, impact, and disruptiveness for each post-Moore technology class, while we discuss in the following paragraphs best practices for the novel technologies adoptions in the simulation methods and plasma use cases investigated in this assessment.\\

\begin{table*}[t!]
\centering
\caption{Best practices for adopting post-Moore technologies in plasma simulations workflows.}
\footnotesize
\setlength{\tabcolsep}{4pt}
\renewcommand{\arraystretch}{1.12}
\begin{tabularx}{\textwidth}{>{\raggedright\arraybackslash}p{2.3cm}>{\raggedright\arraybackslash}p{1.7cm}>{\raggedright\arraybackslash}p{3.7cm}>{\raggedright\arraybackslash}p{3.6cm}>{\raggedright\arraybackslash}X}
\toprule
\textbf{Technology} & \textbf{Readiness} & \textbf{Highest credible impact} & \textbf{Primary plasma workflow target} & \textbf{Disruptiveness} \\
\midrule
FPGA / CGRA & High for diagnostics/control; medium for simulation kernels & High impact on latency-critical pipelines and on narrow PIC/particle-mesh kernels that can be streamed efficiently & Real-time diagnostics and control; PIC interpolation, deposition, sorting, and other locality-controlled subroutines & Low-medium; complementary accelerator rather than a replacement for CPUs/GPUs \\
\midrule
SmartNIC / DPU & Medium-high & Medium impact at workflow level through reduced host overhead, data movement, and in-transit preprocessing & Compression, repartitioning, filtering, staging, and communication-adjacent services around large simulations & Low; incremental workflow optimization with limited impact on solver structure \\
\midrule
Neuromorphic computing & Medium for control/sensing; low-medium for sparse kernels and surrogates & Medium impact where sparse temporal structure and strict latency or power constraints matter & Event-driven diagnostics, control, sparse PDE kernels, and reduced surrogate models & Medium; requires a new programming model, but only for narrow workflow stages \\
\midrule
Thermodynamic computing & Low & Potentially useful acceleration for sampling, Bayesian inference, and matrix-function subroutines in reduced models & Uncertainty quantification, equilibrium reconstruction, Bayesian analysis, and sampling-centric reduced workflows & Medium-high; exploratory and subroutine-focused rather than production ready \\
\midrule
CIM / AIMC & Low-medium & Medium impact when repeated linear algebra dominates and mixed-precision refinement is acceptable & Poisson-like solves, preconditioners, reduced implicit models, and analysis pipelines & Medium; changes the implementation of selected operators without redefining the full workflow \\
\midrule
Photonic computing & Low & Potentially high for narrow operator classes, but low current end-to-end workflow readiness & Structured PDE operators, matrix inversion or MAC-heavy kernels, and transform-heavy surrogate components & Medium; the new hardware will not redefine the full workflow, still immature for production plasma simulations \\
\midrule
Many-body processors & High in adjacent domains; low for plasma-specific adoption & High impact when particle interaction kernels dominate the cost and data path is stable & Strongly coupled plasma MD, WDM microphysics, and Coulomb-dominated particle interactions & High for the affected subfield, not employed community-wide \\
\midrule
Quantum computing & Low & Potentially transformative for selected kinetic demonstrators and WDM/ICF microphysics & Hamiltonian-sim. demonstrators, stopping power, EOS, opacity, and finite-temperature response & Very high; fundamentally disruptive and long-horizon \\
\bottomrule
\end{tabularx}
\label{tab:best_practice}
\end{table*}

\noindent\textbf{PIC and particle-based kinetic methods.}
PIC offers the clearest near-term adoption path because its workflow already separates particle push, particle-mesh coupling, sorting, field solve, and I/O services. The best practice is to keep the end-to-end code on conventional heterogeneous platforms and offload only those stages that can be stabilized into explicit dataflow. In practice, this points to FPGA-class acceleration for interpolation, deposition, and sorting~\cite{Almomany2024FPGA_ElectrostaticPIC,Sakai2016FullPIC_CPUFPGA}, and to SmartNIC/DPU platforms for compression, filtering, repartitioning, and in-transit particle-data handling~\cite{Liu2022SmartNICParticleFlows,Li2024BlueFieldCompression,Suresh2024BlueField3_vecOp_krylov}. This is a medium-readiness, high-impact, and preserves the dominant CPU/GPU programming model. By contrast, end-to-end FPGA migration, neuromorphic reformulation of the full PIC loop, or quantum replacement of production PIC remains premature. Hybrid quantum or surrogate components are so far restricted to reduced demonstrators~\cite{Higuchi2025JPP_HS,hegde2026hybrid_pic}.

\noindent\textbf{Continuum Vlasov and gyrokinetic solvers.}
For continuum kinetic and gyrokinetic codes, the main bottlenecks are usually memory capacity, data movement, repeated operator application, reductions, and communication. Poisson/Maxwell solves, preconditioners, spectral or matrix-function operators, and diagnostic reductions can be treated as replaceable kernels. CIM/AIMC, photonic devices, and thermodynamic computing are the most relevant post-Moore directions for these operator-dominated substeps~\cite{ZidanEtAl2018MemristorPDESolver,Li2023cim_sparse_matrix,Zuo2025cim_matrix_solve,Chen2024photonic_iterative,Tang2025photonic_pde_neural,Aifer2024thermodynamic_linearalgebra,Duffield2025thermodynamics_matrix_exp}, but their readiness is still low-medium to low. SmartNIC/DPU technologies remain relevant for staging, data reduction, and communication-related services, while quantum approaches are confined to linearized or hybrid Hamiltonian demonstrators rather than full nonlinear production workflows~\cite{Engel2019VlasovPRA,Ameri2023VlasovCollisionsPRA,Toyoizumi2024QSVT_HS_PRA}.

\noindent\textbf{Fluid/MHD and Hybrid workflows.}
In this part of the plasma ecosystem, post-Moore technologies matter most when the workflow includes real-time control, inference, or multi-model coupling For hybrid fluid-kinetic workflows, the main recommendation is to make the coupling surfaces modular so that data exchange, field solves, and inference components can be targeted independently by SmartNIC/DPU, FPGA, or future sparse-kernel accelerators rather than attempting a disruptive hardware rewrite of the entire coupled code.

\noindent\textbf{WDM, strongly coupled plasma, and microphysics.}
Warm dense matter and strongly coupled microphysics form the clearest case where more disruptive technologies may eventually matter most. When interaction kernels dominate the cost, the historical lesson from GRAPE, MDGRAPE, and Anton is that specialized many-body hardware can be worthwhile~\cite{Makino1997grape4,Ohmura2014mdgrape4,morimoto2021mdgrape4a,shaw2008Anton,shaw2014anton2}. While for inherent quantum many-body dynamics, quantum computing becomes the strategically long-horizon direction~\cite{vorberger2025roadmap,Rubin2024StoppingPowerPNAS,chen_efficient_2025,rouze_efficient_2025,rouze_optimal_2026}. In this subfield, low current readiness should not be confused with low importance: quantum computing is the least ready technology reviewed here, but potentially the most disruptive.

\section{Conclusion}
\label{sec:conclusions}
This paper has assessed the relevance of emerging post-Moore technologies for plasma simulations from a community and codesign oriented perspective. The analysis indicated that the continued advancement of plasma simulation, particularly for kinetic methods, will depend not only on increased peak performance, but also on the ability to mitigate the growing costs associated with data movement, synchronization, memory access, and communication overhead.

The central finding is that post-Moore technologies are most likely to benefit plasma science by integrating into the parts of the simulation and analysis workflow where their architectural strengths can be exploited most effectively, rather than replacing existing HPC platforms. Among the technologies reviewed, three tiers of maturity and impact emerge. FPGA-class devices and data-path accelerators, including SmartNICs and DPUs, offer the most mature near-term path: FPGAs for latency-critical diagnostics, control, and streamable PIC sub-kernels, and SmartNIC/DPU platforms for workflow-level data services such as compression, filtering, and in-transit analytics. Non-von Neumann architectures, including neuromorphic, thermodynamic, compute-in-memory, photonic, and domain-specific many-body processors, represent medium- to long-term research directions whose current plasma relevance is limited to narrow kernels and proof-of-principle studies, but whose strategic value lies in enabling new algorithms based on novel computing paradigms. Quantum computing, while the least ready technology reviewed, is potentially the most disruptive. Its most coherent role in plasma science targets Hamiltonian-simulation-based treatments of kinetic plasma models, warm dense matter, and inertial confinement fusion microphysics, where structured demonstrator development and resource analysis are the most productive near-term activities.

The immediate priority for the plasma computing community should be to pursue focused demonstrators, benchmark representative plasma kernels and end-to-end workflows on emerging platforms, and strengthen the software and integration ecosystem needed to evaluate these technologies under realistic scientific conditions.

\section*{Acknowledgments}
This work is funded by the European Union. This work has received funding from the European High Performance Computing Joint Undertaking (JU) and Sweden, Finland, Germany, Greece, France, Slovenia, Spain, and the Czech Republic under grant agreement No. 101093261, Plasma-PEPSC.

\bibliographystyle{ACM-Reference-Format}
\bibliography{bibliography}

\end{document}